# Photonic localization probe of molecular specific intranuclear structural alterations in brain cells due to fetal alcoholism via confocal microscopy

Shiva Bhandari,[a] Pradeep K. Shukla,[b] Peeyush Sahay,[c] Avtar S. Meena,[b] Prakash Adhikari,[c] Radhakrishna Rao,[b] Prabhakar Pradhan[c*]

[a] *Department of Physics, East Carolina University, Greenville, NC, USA, 27858*
[b] *Department of Physiology, University of Tennessee Health Science Center, Memphis, TN, USA, 38103*
[c] *Department of Physics and Astronomy, Mississippi State University, Mississippi State, MS, USA, 39762*

*E-mails: PPradhan: pp838@msstate.edu and PShukla: pshukla2@uthsc.edu

Molecular specific photonic localization is a sensitive technique to probe the structural alterations/abnormalities in a cell such as abnormalities due to alcohol or other drugs. Alcohol consumption during pregnancy by mother, or fetal alcoholism, is one of the major factors of mental retardation in children. Fetal alcohol syndrome and alcohol related neurodevelopmental disorder are awful outcomes of the maternal alcohol consumption linked with notable cognitive and behavioral defects. Alcohol consumed by the pregnant mother, being teratogenic, interferes with the fetal health resulting brain damage and other birth defects. This might affect the brain cells at the very nano-level which can't be predicted by the present histopathological procedures. We perform quantification of nanoscale spatial structural alterations in two different spatial molecular components, DNA and histone molecular mass densities, in brain cell nuclei of fetal alcohol effected(FAE) pups at postnatal day 60. Confocal images of the brain cells are collected and the degree of morphological alterations in DNA and histone, in terms of mass density fluctuations are obtained using the recently developed molecular specific light/photonics localization analysis technique. The results show an increase in degree of spatial structural disorder in DNA and a reduced histone modification. Increase in spatial disorder in DNA may suggest DNA unwinding and possibly responsible for increase in gene expression. Reduced histone modification may suggest its release from the DNA and help in the unwinding of DNA and gene expression. The probable cause for structural disorder, as well as opposite rearrangements for DNA and histone molecules in fetal alcohol effects are also discussed.

## 1  Introduction

Light is an important probe to detect structural properties of biological cells. It is now shown that the intracellular structural alterations happen inside the cell in its disease or abnormal state. There are different levels of molecular alterations happening in cell that are related to disease or abnormalities. Molecular specific photonic localization is a sensitive technique to probe and quantify the structural alterations/abnormalities in a cell such as abnormalities due to alcohol or other drugs in brain cells, including alterations in brain cells of new born in fetal alcoholism.

Alcohol has been a part of human culture since the beginning and is responsible for several health issues in the present world. Major health problems reported are directly or indirectly concerned with alcohol. The extent of damage by the alcohol depends on various factors like age, sex, amount and concentration of alcohol, etc. Alcohol affects all the major organs of the body like liver, heart, brain, pancreas, etc. The immune system is degraded by alcohol and the body is prone to disease thereafter [1]. Therefore, alcohol consumption is a cause for increased illness and death since long time [2]. Researchers have shown that complex and multidimensional relationship occur between alcohol consumption and health consequences [3]. Alcohol is, therefore, an important factor responsible for challenging the health issues for the century. Centre for Disease Control and Prevention estimated death toll of 88,000 annually from alcohol related cases and alcohol is ranked as the third leading preventable cause of death in the US. The number of deaths due to alcohol is increasing year by year.

Alcohol has been identified as a carcinogen and is responsible for majority of the cancers [4]. Most of the researches have shown that alcohol has more effect in women than in men [5]. Alcohol consumption by mother during pregnancy is one of the major risk factors of mental retardation of the children born to such mothers in the United States [6]. Alcohol consumption during pregnancy leads to several critical health issues like miscarriage, stillbirth, and different disabilities known as fetal alcohol spectrum disorders (FASDs) [7]. Fetal alcohol syndrome (FAS) and alcohol-related neurodevelopmental disorder (ARND) are the awful outcomes of the maternal alcohol consumption, linked with notable cognitive and behavioral deficits [8]. One child born of every 13 women who consumed alcohol in pregnancy is found to have fetal alcohol disorder syndrome. The number of infants born with FADS is estimated to 630,000 annually worldwide and this number is increasing every year [9]. This issue, which is yet preventable, is tragic as it is a leading cause of intellectual disability, birth defects and developmental disorders [10].

Several risk factors are responsible for analyzing the effects of alcohol exposure on the fetal mental development. The conditions changing the blood alcohol concentration (BAC) in the fetus play important role in influencing the occurrence and harshness of alcohol-induced developmental brain-injuries [11]. The alcohol consumed by the pregnant mother passes through the placenta from her to the growing baby in the womb and has the same amount of alcohol as the mother. The baby is unable to metabolize the alcohol through liver or any other organs and alcohol, being teratogenic, interferes with the healthy development of the baby resulting in brain damages and other birth effects. This might affect the brain cells/tissues at the very nano to submicron level which can't be predicted by the ongoing histological procedures. Therefore, the nanoscale study of the changes in the children brain cell nuclei because of the fetal alcoholism may provide plethora of information.

Most of the changes that occur due to alcohol effect at nano-scale level are not predicted by the existing conventional microscopy. Since these changes are found occurring in the nuclear components and the nuclear components like DNA fall under the diffraction limit of the conventional microscope (i.e. <200nm), therefore, using the confocal technique, we can probe changes around ~150-200nm. For this purpose, we have used light localization based Inverse Participation Ratio analysis using the confocal images to study the effects of this important factor in the brain cells.

## 2  Method

### 2.1 Fetal Alcohol Exposure

All animal experiments were performed according to the protocol approved by the University of Tennessee Health Science Center (UTHSC) Institutional Animal Care and Use Committee. Mice were housed in groups of 2-5 per cage, segregated by sex, in a room on a 12 h/12 h light/dark cycle (lights on at 8:00 AM, off at 8:00 PM) maintained at 22±2 °C. Pregnant mice were fed with or without  (0% 2 days, 1% 2 days, 2% 2 days, 4% 1 week, and 5% 1 week) in Lieber-DeCarli liquid diet (FAE). Control group was pair fed with isocaloric diet. Brains from pair fed (PF) and FAE (EF) offspring were collected at 60-day postnatal age and cryofixed for fluorescence staining.

### 2.2 Florescence Microscopy

The immunofluorescence staining was performed in brain sections with DAPI, the nuclear dye, and immunofluorescence staining of Histone H3K27me3. We have collected and analyzed images from frontal cortex region.  Cryosections (10 μm thick) were fixed in one to one mixture of acetone and methanol at -20°C for 2 minutes and rehydrated in phosphate buffered saline (PBS). PBS is a mixture of 137 mM sodium chloride, 2.7 mM potassium chloride, 10 mM disodium hydrogen phosphate and 1.8 mM potassium dihydrogen phosphate. Sections were permeabilized with 0.2% Triton X-100 in PBS for 10 minutes and blocked in 4% non-fat milk in Triton–Tris buffer (150 mM sodium chloride containing 10% Tween-20 and 20 mM Tris, pH 7.4). It was then incubated for 1 hour with the primary antibodies (rabbit polyclonal anti-H3k27me3), followed by incubation for 1 hour with secondary antibodies (cy3-conjugated anti-rabbit IgG antibodies). The fluorescence was examined using a Zeiss710 confocal microscope (Carl Zeiss Microscopy, Jena, Germany), and images from x–y sections (1 lm) were collected using LSM 5 Pascal software (Carl Zeiss Microscopy). Images were stacked by using the software ImageJ (NIH, Bethesda, MD), and processed by Adobe Photoshop (Adobe Systems Inc., San Jose, CA). All images for tissue samples from different group were collected and processed under identical conditions. The images obtained were categorized into four groups as (i) PF-DAPI stained brain cells, (ii) EF-DAPI stained brain cells, (iii) PF-H3K27me3 stained brain cells, and (iv) EF-H3K27me3-stained brain cells.

**2.3 Molecular specific confocal imaging and intracellular structural disorder analysis**

The method of confocal imaging and quantification of light localization properties of the biological system has been described previously and the details is given elsewhere [12,13]. In brief, confocal microscopy is a powerful optical imaging technique that increases the optical resolution and contrast of the images. It uses a spatial pinhole for blocking the out-of-focus light during the image formation. The fluorescence emitted by the sample molecules from small finite volume or voxel *dV (=dxdydz)* around the excitation center is collected by the microscope objective and imaged onto the detection pinhole in front of the photo detector. The confocal image intensity is found from previous study [13] to be:

$$I(r) \propto dV(\rho), \qquad (1)$$

Where *I(r)* is the pixel intensity of the confocal image at position *r*, $\rho$ is the density of the molecules at small volume *dV* (or, of the voxel *dV= dxdydz*) in the sample.

It has been shown that the local refractive index of a cell is proportional to its local mass density [13-15], i.e. *n(r) =$n_o$ + Δn(r)*. The pixel intensity values of the confocal images can be correlated to the refractive index fluctuations in the corresponding fluorescence molecules mass density variation. So, a representative refractive index matrix can be constructed using the pixel intensity values, with the optical potential $\mathcal{E}(x, y)$ defined as:

$$\varepsilon(r) = \frac{\Delta n(x, y)}{n_0} \propto \frac{dI(x, y)}{<I>} \quad . \tag{2}$$

Where $n_0$ and $\Delta n(x, y)$ denote the average refractive index of the fluorescent molecules and its fluctuation at $(x, y)$ position (x-y size dxdy and sample thickness $dz$), respectively. $<I>$ is the average intensity of the confocal images and $dI(x, y)$ is the fluctuation in the intensity at $(x, y)$ pixel position of the confocal image. The corresponding pixel intensity values in the confocal image thus, give the onsite optical potential, $\mathcal{E}(x,y)$. From this, it is clear that an optical lattice is a representation of the spatial refractive index fluctuations of the fluorescent molecules inside the sample [16].

The spatial structural disorder strength of an optical lattice, produced from the confocal image, can be analyzed using Anderson Tight Binding Model (TBM) Hamiltonian approach. Anderson TBM is well studied in the condensed matter physics for describing the disorder properties of optical systems of any geometry and disorder [17]. If we consider one optical state per lattice site, with inter lattice site hopping restricted to the nearest neighbors only, the tight binding Hamiltonian can be written as:

$$H = \sum_i \varepsilon_i |i\rangle\langle i| + t \sum_{\langle ij \rangle} (|i\rangle\langle j| + |j\rangle\langle i|). \tag{3}$$

Where $\mathcal{E}_i$ is the lattice optical potential corresponding to the $i^{th}$ lattice site, $j$ is the nearest neighbor of $i^{th}$ lattice site and $t$ is the inter-lattice site hopping strength.

The eigenfunctions and eigenvalues are obtained from the Hamiltonian. The average IPR value for a lattice system is calculated from the above Hamiltonian as

$$\langle IPR \rangle = \frac{1}{N} \sum_{i=1}^{N} \int_0^L \int_0^L E_i^4(x, y)\, dxdy \tag{4}$$

Where $E_i$ is the i$^{th}$ eigenfunction of the Hamiltonian of optical lattice of size $L \times L$, having N lattice point.

Biological cell systems are heterogenous light transparent media and hence can be represented as refractive index systems. Therefore, the disorder in these biological systems can be specified by two parameters, refractive index fluctuations $\Delta n$ and its spatial fluctuation correlation length $l_c$. These two parameters combined together provides the measure of the refractive index fluctuations inside the system and is termed as disorder strength, $L_d$:

$$L_d = \langle \Delta n \rangle \times l_c \qquad (5)$$

and the average value of *IPR* represents the measure of the disorder strength, so IPR and disorder strength can be written as

$$\langle IPR \rangle \propto L_d = \langle \Delta n \rangle \times l_c \; . \qquad (6)$$

## 3  Results

We have applied IPR technique to study the effect of alcohol in brain cell nuclei of FAE pups using the brain cell nuclei confocal images. The confocal images were obtained from the FAE pups' brain cells collected at the postnatal day 60. The <IPR> values of the confocal images from four different categories as indicated in Method 2.2 are obtained.

This IPR technique is used to analyze the morphological changes in the pup's brain cells due to fetal alcohol exposure. We analyze the changes that occurred in DNA and histone protein of the fetal ethanol exposed pup's brain. For this, we perform a comparative study between (i) control fed (PF) DAPI stained brain cells and ethanol fed (EF) DAPI stained brain cells, and (ii) control fed (PF) H3K27me3 stained brain cells and ethanol fed (EF) H3K27me3 stained brain cells. We performed IPR imaging and analysis for all the four categories of the samples under study.

*Results of DNA molecular specific DAPI staining:*

Figures 1(a) and 1(b) show the DAPI stained brain cells for control fed and ethanol fed respectively, and figures 1(a') and 1(b') show the corresponding <IPR> or structural disorder of these cells. DAPI staining mainly target DNA molecules. Figure 2 shows the bar plot for the

…

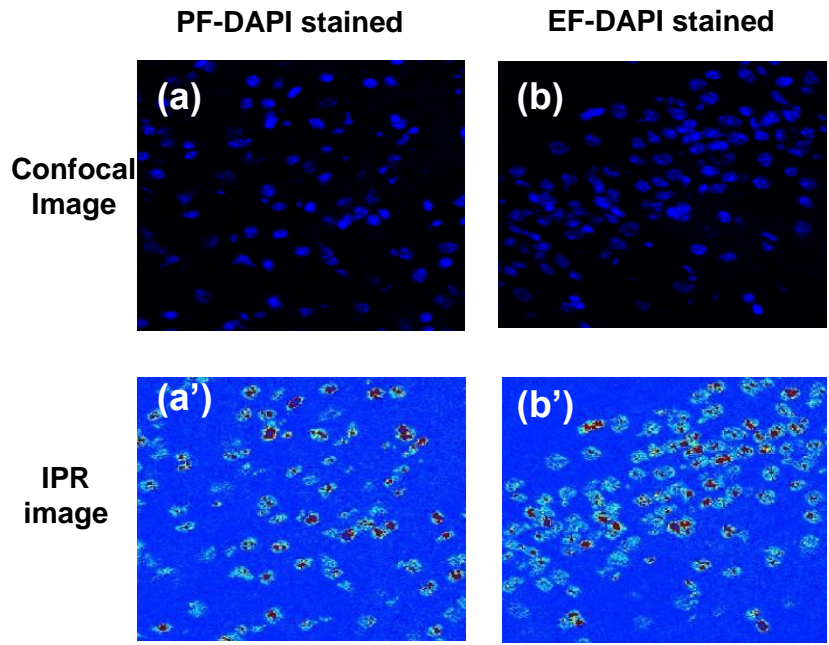

**Fig.1: (a)** and **(b)** represent the confocal images of the control fed (PF) DAPI stained brain cells and ethanol fed (EF) DAPI stained brain cells respectively. **(a')** and **(b')** represent the IPR images of the control fed (PF) DAPI stained brain cells and ethanol fed (EF) DAPI stained brain cells respectively.

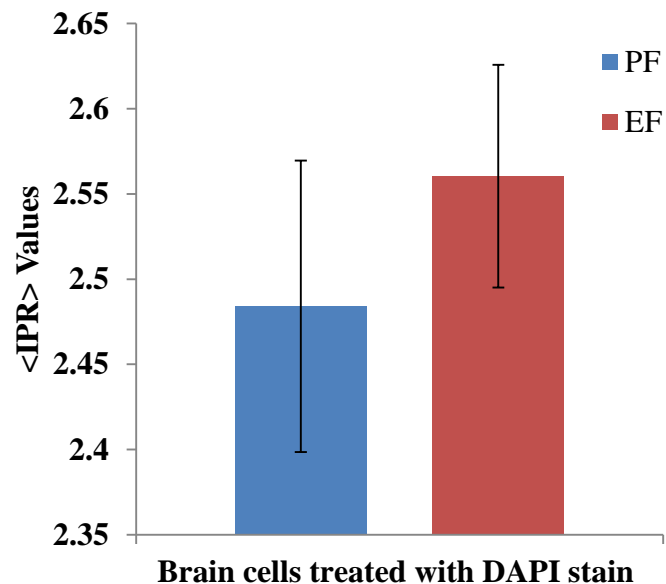

**Fig 2:** *<IPR>* values for control fed (PF) DAPI stained brain cells and ethanol fed (EF) DAPI stained brain cells.-The IPR value for ethanol fed DAPI stained brain cells is found higher as compared to that of control fed DAPI stained brain cells. As DAPI binds with DNA molecules, the above graph shows that fetal ethanol fed pups have more disorder in their brain cell nuclei leading to different kind of abnormalities. The percentage difference of the *<IPR>* values between these two groups is found to be 6%. (P-value <0.05, n=10).

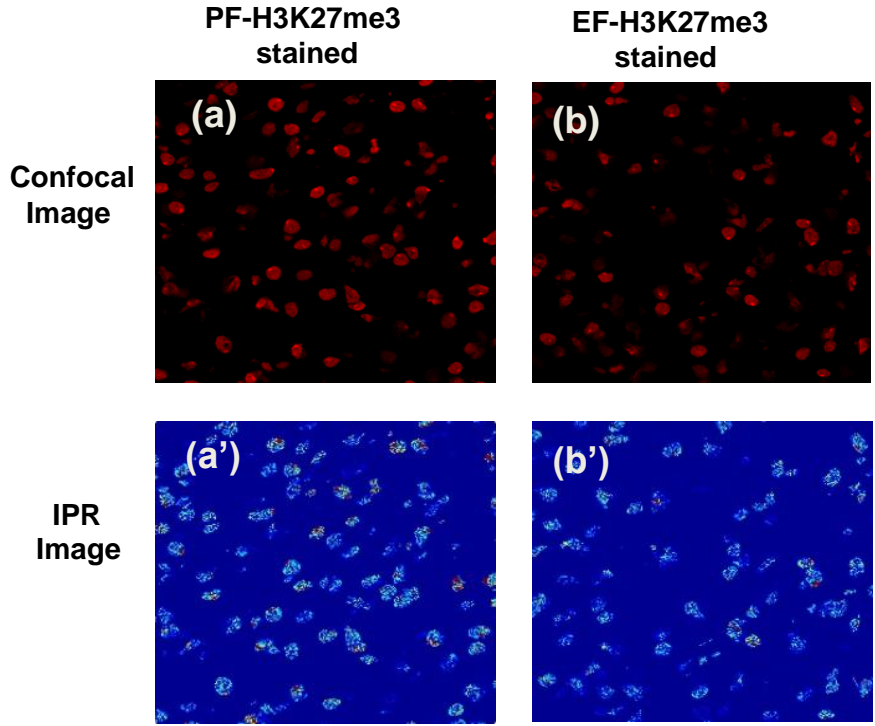

**Fig.3:** **(a)** and **(b)** represent the confocal images of the control fed (PF) H3K27me3 stained brain cells and ethanol fed (EF) H3K27me3 stained brain cells respectively, targeting histone. **(c')** and **(d')** represent the IPR images of the control fed (PF) H3K27me3 stained brain cells and ethanol fed (EF) H3K27me3 stained brain cells respectively.

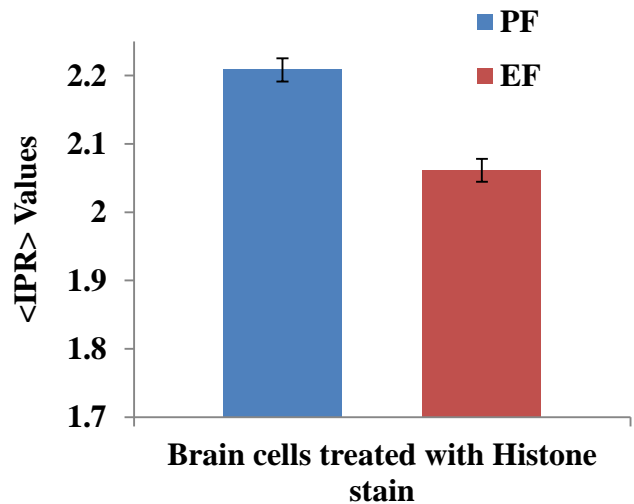

**Fig. 4**: *<IPR>* values or the degree of structural disorder for control fed (PF) H3K27me3 stained brain cells and ethanol fed (EF) H3K27me3 stained brain cells. The *<IPR>* value for ethanol fed (PF) H3K27me3 stained brain cells is found lower as compared to that of the control fed (EF) H3K27me3 stained brain cells. The percentage difference of the *<IPR>* values for structural disorder between these two groups is found to be 7 %. (P-value < 0.05, n=10).

comparisons of <IPR> values for control fed (PF) DAPI stained brain cells and ethanol fed (EF) DAPI stained brain cells.

*Results of histone molecular specific H3K27me3 staining:*

Figures 3(a) and 3(b) show H3K27me3 stained brain cells for the control fed and ethanol fed brain cells respectively. Figures 3(a') and 3(b') show the corresponding IPR, or structural disorder strength. H3K27me3 staining mainly target the histone molecules in the brain cell nuclei. Figure 4 shows the bar graph comparisons of ensemble averaged *<IPR>* values for control fed (PF) H3K27me3 stained brain cells and ethanol fed (EF) H3K27me3 stained brain cells. All these *<IPR>* analysis were done at a sample size ~4 μm. The graphs show the trend of changes occurring in histone protein due to fetal alcoholism. Histone structures in brain cells exposed to fetal alcoholism show less structural disorder relative to the control. This is opposite to the nuclear DNA molecular structure, which show the increase in structural disorder for exposition to the fetal alcoholism.

Many studies have shown that there are changes in the DNA and histone proteins due to alcohol consumption, but there is no clear understanding of the level of changes due to alcohol. From our study, we show that there are structural changes occurring at the DNA and histone protein due to the alcohol. But the changes occurring in these two components of chromatin, i.e., DNA and histone, are opposite, in particular the degree of structural disorder increase in DNA and decrease in histone due to the fetal alcoholism.

The IPR values for DNA molecules of ethanol fed pup's brain cell nuclei are found to be greater than the control one. IPR value is correlated with the structural disorder, so that the spatial distribution of DNA molecules in ethanol fed pup's brain cells have more structural disorder. So, it can be suggested that the introduction of alcohol results in the spatial variation in the components of DNA and hence, responsible for the changes in the genetics to introduce different fetal alcoholic syndrome.

The <IPR> value of the ethanol fed H3K27me3 stained brain cells is found to be lesser than the IPR value of the control ethanol fed H3K27me3 stained brain cells. This suggests that the change in the degree of structural disorder strength of the ethanol fed H3K27me3 stained brain cells is less than the control one. This means, less mass density fluctuations in the histone proteins in brain cells due to fetal alcoholic exposition. So, there could be suppression in certain gene expression resulting the fetal alcoholic syndrome disorder in the brain histone molecules.

# 4  Conclusion and Discussion

The effect of chronic alcoholism during the pregnancy on the fetus, or the effect of fetal alcoholism is studied using fetal alcoholic mice model. In particular, we studied spatial structural properties of molecular mass density variations of two types of molecules: i) DNA molecular mass density and ii) histone protein molecular mass density. Fetal alcohol exposure (FAE) seems to affect the DNA and histone protein molecules in the brain cell nuclei of ethanol fed pups in reverse way. We perform the analysis on the confocal images of the brain cells of ethanol fed pups and control fed pups at two different conditions. Alcohol is responsible for the structural changes in DNA and histone proteins. The changes in these two components of chromatin are opposite. Alcohol is responsible for the DNA methylation process. DNA methylation results into the compact nucleosome structure which could be responsible for greater mass fluctuations in the DNA and hence greater *<IPR>* value. The higher <IPR> value of ethanol fed DAPI stained pup's brain cell nuclei suggests that the introduction of alcohol results in the spatial variation in the components of DNA. This spatial variation induces the changes in the genetics to introduce different fetal alcoholic syndrome.

*Probable reasons for opposite changes in DNA and histone molecules in fetal alcoholism:* The fetal alcohol exposure might have caused the histone protein modifications and these histone protein modifications could be responsible for enhancing the relaxation of chromatin causing less mass fluctuations and hence smaller value of *<IPR>*. H3K27me3 denotes the trimethylation of lysine 27 on histone protein. Addition of alcohol causes the loss of H3K27me3 methylation process. This loss of methylation may play a role in DNA unwinding and gene expression. The lower value of *<IPR>* is associated with low disorder. This disorder may be due to the loss of methylation occurring in the loosely attached histones but not in the histones that are tightly attached to the DNA. This may result in the loose packing of nucleosomes and hence the relaxed state of chromatin. The relaxed state of chromatin may be the cause of small disorder. Because of these changes, there could be upregulation or downregulation in certain gene expression resulting the fetal alcoholic syndrome disorder in the brain. Further exploration on the fetal alcohol syndrome due to the structural alterations in the DNA and histone is needed. This result could be a basis for the further study.

**Acknowledgements:**

The work reported here was partially supported by the National Institutes of Health (NIH) grants (R01EB003682 and R01EB016983) and Mississippi State University to Dr. Pradhan. Dr. Rao was supported by NIH grants (AA12307 and DK55532). SV and PP also thank UofM , where part of this work was completed.